\documentclass[aps,pre,amsmath,amsfonts,amssymb,superscriptaddress,showpacs,10pt]{revtex4}
\usepackage{graphics}
\usepackage[latin1]{inputenc}
\usepackage{amscd}

\usepackage[pdftex]{graphicx}

\usepackage{rotating}

\usepackage{color}

\usepackage{amsthm}
\usepackage{framed}
\usepackage{amssymb}
\usepackage{amsmath,mathrsfs}
\usepackage{hhline}
\usepackage{fancyhdr}
\usepackage{tikz}
\everymath{\displaystyle}
\everymath{\displaystyle}

\everymath{\displaystyle}
\newcommand{\beq}{\begin{equation}}
\newcommand{\eeq}{\end{equation}}

\newcommand{\bi}{\begin{itemize}}
\newcommand{\ei}{\end{itemize}}

\def\RR{\mathbb{R}}
\def\ZZ{\mathbb{Z}}
\def\NN{\mathbb{N}}

\begin{document}
\title{Catching homologies by geometric entropy}

\author{Domenico Felice}
\email{domenico.felice@unicam.it}
\affiliation{School of Science and Technology,
University of Camerino, I-62032 Camerino, Italy \\
INFN-Sezione di Perugia, Via A. Pascoli, I-06123 Perugia, Italy}
\author{ Roberto Franzosi}
\email{roberto.franzosi@ino.it}
\affiliation{QSTAR and INO-CNR, largo Enrico Fermi 2, I-50125 Firenze, Italy}
\author{Stefano Mancini}
\email{stefano.mancini@unicam.it}
\affiliation{School of Science and Technology,
University of Camerino, I-62032 Camerino, Italy \\
INFN-Sezione di Perugia, Via A. Pascoli, I-06123 Perugia, Italy}
\author{Marco Pettini}
\email{pettini@cpt.univ-mrs.fr}
\affiliation{ Aix-Marseille University, Marseille, France\\
CNRS Centre de Physique Th\'eorique UMR7332,
13288 Marseille, France}

\begin{abstract}
A geometric entropy is defined as the Riemannian volume of the parameter space of a statistical manifold associated with a given network. As such it can be a good candidate for measuring networks complexity. Here we investigate its ability to single out topological features of networks proceeding in a bottom-up manner: first we consider small size networks by analytical methods and then large size networks by numerical techniques. Two different classes of networks, the random graphs and the scale--free networks, are investigated computing their Betti numbers and then showing the capability of geometric entropy of detecting homologies.
\end{abstract}


\maketitle

\section{Introduction}

Common understanding identifies a \textit{network} as a set of items, called \textit{nodes} (or vertices), with connections between them, called \textit{links} (or edges) \cite{Newman}. Since many systems in the real word take the form of networks (also called \textit{graphs} in much of the mathematical literature), they are extensively studied in many branches of science, like, for instance, social, technological, biological and physical science \cite{Boccaletti06}. At the beginning, the study of networks was one of the fundamental topics in discrete mathematics: Euler is ascribed as the first providing a true proof in the theory of networks by its solution of the K\"onigsberg problem in 1735. Recently, also thanks to the availability of computers, the study of networks moved from the analysis of single small graphs and the properties of individual nodes and links within such graphs to the consideration of large-scale statistical properties of graphs. Thus, statistical methods became a prominent tool to quantify the degree of organization (\textit{complexity}) of large networks \cite{barabasi1}.

A typical approach in statistical mechanics of complex networks is the statistical ensemble. Such an approach is a natural extension of Erd\"os-R\'enyi ideas \cite{ER}. It has been performed through two basic ideas: the configuration space weight and the functional weight \cite{BBW06}. The first one is proportional to the uniform probability measure on the configuration space which accounts for the way to uniformly chose graphs in the configuration space. Whereas,  functional weights depend on the network topologies and are chosen in order to address the statistical mechanics approach to networks different from the random graphs, which have some typical structures, like the \textit{small world} property \cite{storogatz}, the power-law degree distribution \cite{barabasi2}, the correlation of node degrees \cite{bollobas}, to name the most frequently addressed. During the last decade, several works have been inspired by this approach \cite{Bianconi}.

Recently, the techniques of statistical mechanics were complemented by new topological methods: a network is encoded through a simplicial complex which can be considered as a combinatorial version of a topological space whose properties can be studied from combinatorial, topological or algebraic points of view. Thus, regarding the mentioned topological aspects, different measures of simplicial complexes and of networks stemming from simplicial complexes can be defined. This provides a link between topological properties of simplicial complexes and statistical mechanics of networks from which simplicial complexes were constructed \cite{jstat}.

In this work we consider a geometric entropy which is inspired by microcanonical entropy of statistical mechanics and stems from Information Geometry (IG) \cite{FMP}. In particular, a Riemannian manifold (differentiable object) is associated to a network (discrete object) and the complexity measure is the logarithm of the volume of the manifold. More precisely,  random variables are associated to each node of a network and their correlations are considered as weighted links among the nodes. The nature of these variables characterizes the network (for example, each node can contain energy, or information, or represent some internal parameter of a neuron in a neural network, or the concentration of a biomolecule in a complex network of biochemical reactions, and so on). The variables are assumed to be random either because of the difficulty of perfectly knowing their values or because of their intrinsic random dynamical properties. Thus, as it is customary for probabilistic graphs models \cite{keshav}, a joint probability mass function is associated to the description of the network. At this point, we assume Gaussian joint probability mass functions because of their tractability and since they are used extensively in many applications ranging from neural networks, to wireless communication, from proteins to electronic circuits, etc. Finally, the geometric  complexity measure of networks is obtained  by resorting to the afore mentioned relation between networks and joint probability mass functions and introducing in the space of these mass functions  a Riemannian structure borrowed from information geometry \cite{amari}.

In addition to the statistical methods in network complexity we also consider topological methods by encoding a network into a \textit{clique graph} $C(G)$, that has the complete subgraphs as simplexes and the nodes of the graph (network) $G$ as its nodes so that it is essentially the complete subgraph complex. The maximal simplexes are given by the collection of nodes that make up the cliques of $G$. In particular, we are interested in the information about the topological space $C(G)$ stored in the number and type of holes it contains. So, we exploit algebraic topology tools in order to describe a network by means of its homology groups \cite{Carlsson}. 

The ability of the geometric entropy to capture algebraic topological features of networks is investigated by a bottom-up approach. First, we consider networks with low number of nodes (\textit{small--size}) and given homology groups. Then, we compute the dimension of the homology groups of large--size networks. Hence, we compare the geometric entropy against the dimensions of homology groups (Betti numbers) revealing a clear detection of topological properties of the considered networks. In particular, when dealing with large--size networks, we consider random graphs and scale--free networks. According to the well--known transition in the appearance of a giant component\cite{ER,aiello2001}, a description of networks through their Betti numbers shows a clear correlation with the growth of the size of the largest components. This perfectly matches the behaviour of the geometric entropy \cite{Franzosi16} which in turn, when compared to  Betti numbers of the the networks, clearly appears to probe relevant topological aspects of the networks.

The organization of the present paper is as follows. In Section \ref{sec2} we review some methods of the Algebraic Topology useful to describe topological properties of networks encoded in simplicial complexes. In Section \ref{sec3} we describe the geometric entropy stemming from both statistical methods and IG methods. In Section \ref{sec4} we compute the Homology groups of small--size networks as well as of large size networks within the ensembles of random graphs and scale--free networks. Then we compare the geometric entropy computed on these networks against their Betti numbers. Concluding remarks are given in Section \ref{sec5}.

\section{Basics of Algebraic Topology}\label{sec2}

In order to make the present work self contained, we start by reviewing some methods of combinatorial algebraic topology {by referring to} \cite{spanier}; these methods allow a topological characterization of networks. In particular, we focus on simplicial algebraic invariants; among them, we select homologies since they are easier to compute than, for example, {homotopy groups.} 

\subsection{Simplicial Complexes}

Let $I=\{v_i\}_{i\in\NN}$ be a set of vertices (or nodes). A \textit{simplex} $s$ in $I$ with dimension $n$ is any its subset with cardinality equal to $n+1$, and it is called a $n$-simplex; in particular the \textit{empty set} is the only $-1$-simplex. A \textit{face} of a $n$-simplex $s:=\{v_0,v_1,\ldots,v_n\}$ is the simplex $s^\prime$ {whose} vertices consist of any nonempty subset of the $v_i$s; if $s^\prime$ is a $p$-simplex, with $p<n$, it is called a $p$-face of $s$. The subset needs not be a proper subset, so $s=\{v_0,v_1,\ldots,v_n\}$ is regarded as a face of itself. 

\begin{figure}\centering
\vspace{0.2cm}
\begin{tabular}{c}
 \begin{tikzpicture}[thick,scale=0.8]
     \draw[fill=yellow!20] \foreach \x in {0}
     {
          (\x:2) node{0} -- (\x+72:2) node{1} -- (\x+144:2) node{2} -- (\x+216:2) node{3} -- (\x+288:2) node{4}  -- (\x+360:2) node{0}
       (\x:2) -- (\x+144:2) 
       (\x:2) -- (\x+216:2)
     };     
 \end{tikzpicture}
 \end{tabular}\hspace{2.5cm}
\begin{tabular}{c}
 \begin{tikzpicture}[thick,scale=0.8]
     \draw\foreach \x in {0}
     {
         (\x:2) node{0} -- (\x+72:2) node{1} -- (\x+144:2) node{2} -- (\x+216:2) node{3} -- (\x+288:2) node{4}  -- (\x+360:2) node{0}
           };
 \end{tikzpicture}
 \end{tabular}
 \caption{(Left)A polygonal disk with $5$ nodes. (Right)A standard polygon with $5$ edges.} \label{polygon}
 \end{figure}
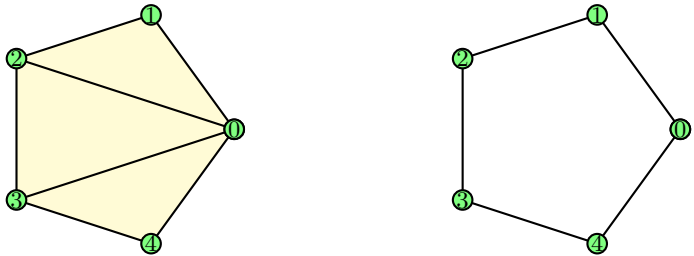

A \textit{simplicial complex} $K$ consists of a set $I$ of vertices and a set $\{s\}$ of simplexes such that 
(i) any set consisting of exactly one vertex is a simplex; (ii) any nonempty subset of a simplex is a simplex.
It follows from condition (i) that $0$-simplexes of $K$ correspond bijectively to vertices of $K$. {Analogously, from condition (ii) it follows} that any simplex is determined by its $0$-faces. Thus, we can identify $K$ {as the set of its simplexes, and a vertex of $K$ as} the $0$-simplex corresponding to it. For example, let $I=\ZZ/{n\ZZ}$ be the set of vertices, and consider the simplicial complex $P_n$ on $I$ with set of simplexes $\{i,i+1\},\,i\in I$. Intuitively, $P_n$ is a set of $n$ vertices and $n$ links (edges) among them. Therefore, $P_n$ is called the \textit{standard polygon} with $n$ edges. If, we add to it the $2$-simplexes $\{0,i,i+1\}$, for $i=1,\ldots,n-2$, we arrive at the simplicial complex $D_n$ called the \textit{standard polygonal disk}. Less formally, in order to obtain $D_n$ we add to the standard polygon $P_n$ the triangles that we can construct among triplets of the set of $n$ vertices. As an example, we draw the difference between a polygon and a polygonal disk in Figure \ref{polygon} when $n=5$.

\tikzstyle{every node}=[circle, draw, fill=green!50,
                         inner sep=0pt, minimum width=4pt]

 The \textit{dimension} of a simplicial complex 
 $K$ {is $\mbox{dim} K=\sup \{\mbox{dim}\, s;s\in K\}$.
 $K$ is said to be \mbox{finite} if it contains only a finite number of simplexes. In such a case} 
 $\mbox{dim} K<\infty$; however, if $\mbox{dim} K<\infty$, $K$ {needs not be finite.} Indeed, consider the simplicial complex $K$ with set of vertices $\ZZ$, and as the set of simplexes the edges $\{i,i+1\}$ as $i$ varies in $\ZZ$; then $K=\{\{i\};i\in\ZZ\}\cup\{\{i,i+1\};i\in\ZZ\}$. In this case, $K$ is not finite but $\mbox{dim} K=1$. A simplicial complex $K$, that  we have abstractly defined upon a set of vertices $I=\{v_i\}_{i=1}^k$, can be also assigned by means of a {\it geometric realisation} by assuming that its vertices are points in $\RR^n$. For example, we get what is known as the natural realisation if we take $n = k + 1$ and $v_0 = e_1, v_1 = e_2,\ldots, v_k = e_{k+1}$, where the $e_i$ are the standard basis vectors in $\RR ^n$. We note that although $K$ may be $n$-dimensional, a realisation $K$ may not ``fit'' into $\RR^n$. Indeed, the standard polygon on $5$ vertices cannot be embedded into $\RR ^1$ even though its dimension amounts to $1$. For this reason, being networks abstract discrete objects, we rely on the combinatorial {approach avoiding any particular geometric realisation.}
 
{When a simplicial complex $K$ is finite, it is possible to introduce the \textit{Euler-Poincar\'e characteristic} $\chi(K)$ as the summation along the number of all its simplexes }
\begin{equation}
 \label{E-P}
 \chi(K)=1+(-\nu_{-1}+\nu_0-\nu_1+\nu_2-\ldots)=1+\sum_{p\in\ZZ}(-1)^p\,\nu_p,
 \end{equation}
 where $\nu_p$ is the number of $p$-simplexes. In particular, we have that $\chi(\emptyset)=0$.
 
 \subsection{Paths and fundamental group}
 
 Let $I$ be a set of vertices, a \textit{step} into a simplicial complex $K$ on $I$ is an element $q=(v,w)\in I\times I$ such that $\{v,w\}\in K$; $v$ is the \textit{initial point} and $w$ is the \textit{final point} of the step. Two steps $q_1,\,q_2$ are \textit{consecutive} if the final point of $q_1$ is the initial point of $q_2$. A \textit{path} is a sequence $\gamma=(q_1,\ldots,q_n)$ of consecutive steps. Two paths $\alpha=(\alpha_1,\ldots,\alpha_r)$ and $\beta=(\beta_1,\ldots,\beta_s)$ can be multiplied if $(\alpha_1,\ldots,\alpha_r,\beta_1,\ldots,\beta_s)$ is a path, i.e. $\alpha_r$ and $\beta_1$ are consecutive; such a path is called the product $\alpha*\beta$.  
 
 Consider $v,w,z\in I$ such that $\{v,w,z\}\in K$; we say that the path $((v,w),(w,z))$ is elementary contractible in the step $(v,z)$. By referring to the {Fig.\ref{polygon}-left the path $((2,1),(1,0))$ can be contracted to the path $(2,0)$, whereas there is no way to contract any paths in  Fig.\ref{polygon}-right.} In general, if a path $\alpha$ has two consecutive steps $((v,w),(w,z))$, we can substitute them with the step $(v,z)$; in this way we obtain a new path $\alpha^\prime$, and we say that $\alpha^\prime$ is obtained from $\alpha$ by an \textit{elementary contraction}. Vice versa, we say that $\alpha$ is obtained from $\alpha^\prime$ by an \textit{elementary expansion}.
 
 Two paths $\alpha$ and $\beta$ in $K$ are called \textit{homotopic} if it is possible to go from one to the other by means of a finite number of elementary expansions or contractions; in this case we write $\alpha\sim\beta$. The latter is an equivalence relation that preserves the initial and the final points; moreover the product ``$*$" passes to the quotient, meaning that if $\alpha\sim\alpha^\prime$ and $\beta\sim\beta^\prime$, then $\alpha*\beta\sim\alpha^\prime*\beta^\prime$. Consider now $i_0\in I$; a \textit{loop} with base point $i_0$ is a path in $K$ which starts from $i_0$ and ends at $i_0$. The set of the homotopy {classes of loops at $i_0$ forms a group $\pi_1(K,i_0)$ under multiplication $*$,} called the \textit{fundamental group} of $K$ at $i_0$.
  
 \subsection{Chains, cycles and boundaries}
  
Given the usual set $I$ of vertices, an \textit{oriented} $p$-simplex of the simplicial complex $K$ on $I$ is a $(p+1)$-tuple $(v_0,v_1,\ldots,v_p)\in I^{p+1}$ such that $\{v_0,v_1,\ldots,v_p\}\in K$. For $p<0$ there are no oriented $p$-simplexes; for every vertex $v\in K$ there is a unique oriented $0$-simplex $(v)$. Then we define the $p$th module $S_p(K,\ZZ)$ of chains in $K$ as the free $\ZZ$-module generated by oriented $p$-simplexes. A $p$-chain is basically a linear combination of $p$-simplexes with coefficients in $\ZZ$. Then it straightforwardly follows that $S_p(K,\ZZ)=0$ for $p<0$. 

Consider now the module--homomorphisms $\partial_p:S_p(K,\ZZ)\rightarrow S_{p-1}(K,\ZZ)$ for $p\geq 1$. Their values are uniquely determined on the $p$-simplexes. So, we can define
\begin{equation}
\label{hom}
\partial_p(v_0,v_1,\ldots,v_p)=\sum_{i=0}^p (-1)^i(v_0,\ldots,\hat{v_i},\ldots,v_p),
\end{equation}
where $(v_0,\ldots,\hat{v_i},\ldots,v_p)$ means the oriented $(p-1)$-simplex obtained by omitting $v_i$. {Actually,} the homomorphism $\partial_p$ is a boundary operator in the sense that it acts on a $p$--simplex by giving rise to its faces. It is not difficult to show that $\partial_p\partial_{p+1}=0$ 
{for all $p$.}

Let $w$ be an oriented $p$-chain in $K$, i.e. $w\in S_p(K,\ZZ)$; then, $w$ is called a $p$-\textit{cycle} if $\partial_p(w)=0$; $w$ is called a $p$-\textit{boundary} if there exists $z\in S_{p+1}(K)$ such that $w=\partial_{p+1}(z)$. The set of the $p$-cycles $Z_p(K,\ZZ)$, and the set of the $p$-boundaries $B_p(K,\ZZ)$ are submodules of $S_p(K,\ZZ)$, {and the relation
$B_p(K,\ZZ)\subset Z_p(K,\ZZ)$ holds true for $p\in\ZZ$.} Less formally, a cycle is a member of $B_p(K,\ZZ)$ if it ``bounds'' something contained in the simplicial complex $K$. For example, by referring to the polygonal disk $D_n$ of Fig. \ref{polygon}, we can see that the oriented chain $(1,0)+(0,2)+(2,1)$ is a boundary, while the chain $(1,0)+(0,4)+(4,3)$ is not.
 
\subsection{Homology groups}
 
Since $B_p(K,\ZZ)\subset Z_p(K,\ZZ)$ for all $p\in\ZZ$, we can consider the quotient module $H_p(K,\ZZ)=\ker \partial_p/\mbox{im}\,\partial_{p+1}=Z_p/B_p$ (called $p$th-module of homology). Intuitively, the construction of homology assumes that we are removing the cycles that are boundaries of higher dimension from the set of all $p$-cycles, so that the ones that remain carry information about $p$-dimensional holes of the simplicial complex. 

If $H_p(K,\ZZ)$ is finitely generated (which is necessarily true if $K$ has finitely many simplexes) from the structure theorem {\cite{spanier}} it follows the $H_p(K,\ZZ)$ is isomorphic to the direct sum of a finite free $\ZZ$-module $H_p^\prime$ and a finite number of finite cyclic groups $\ZZ/n_1\ZZ\oplus\ZZ/n_2\ZZ\oplus\ldots\oplus\ZZ/n_\kappa\ZZ$, where $n_i$ divides $n_{i+1}$. Thus, the $\mbox{rank}(H_p(K,\ZZ))$ is defined as the number of basis elements of $H^\prime$ on $\ZZ$.
{Such a rank is also the $p${th} Betti number of $K$, i.e. $\beta_p:=\mbox{rank}(H_p(K,\ZZ))$,
hence the Euler-Poincar\'e characteristic {\eqref{E-P}} becomes} $\chi(H(K))=1+\sum_{p\in\ZZ}(-1)^p\,\beta_p$.

Let us now compute the lower homology groups. {Consider first an empty simplicial complex $K$}. In this case, it has only one $(-1)$-simplex, the empty set $\emptyset$; thus, the complex chain $S_{-1}(K,\ZZ)$ is isomorphic to $\ZZ$. In addition, $S_p(K,\ZZ)=(0)$ for $p\neq -1$. For these reasons $H_{-1}(\emptyset,\ZZ)\cong \ZZ$ and $H_p(K,\ZZ)=(0)$ for $p\neq -1$. {For non empty $K$ the homomorphism $\partial_0:S_0(K,\ZZ)\rightarrow S_{-1}(K,\ZZ)$ is surjective and $H_{-1}(K,\ZZ)=(0)$. Concerning $H_0(K,\ZZ)$, notice} that
$$\partial_0:S_0(K,\ZZ)\rightarrow S_{-1}(K,\ZZ)(\cong\ZZ)$$
is defined by 
\begin{equation}
\label{boundaryoperator}
\partial_0(r_1\,i_1+r_2\,i_2+\ldots+r_N\,i_N)=r_1+\ldots+r_N,
\end{equation}
for any $i_1,i_2,\ldots,i_N\in\ZZ$. From \eqref{boundaryoperator} and the relation $\partial_0\partial_1\equiv 0$ it follows that, if $(i_0,i_1)$ is a $1$-simplex in $K$, then $\partial_0(i_0,i_1)=i_1-i_0$. For this reason $\partial_0(\sigma)=v_1-v_0$ when $\sigma$ is a path in $K$ from $v_0$ to $v_1$. 
Let us now assume that $K$ is connected and $v_0\in K$. Then, every $0$-cycle $r_1\,i_1+r_2\,i_2+\ldots+r_N\,i_N$ can be written as $r_1\,(i_1-v_0)+r_2\,(i_2-v_0)+\ldots+r_N\,(i_N-v_0)$ and it follows that $H_0(K,\ZZ)=(0)$. {In contrast,} if $K$ is not connected and $V$ is the set of vertices one for any connected component, we have that $H_0(K,\ZZ)$ is the $\ZZ$-free module on the classes $v-v_0$ for $v\in V$. Hence, the notion of connectivity in $K$ is reflected on $H_0(K,\ZZ)$, the dimension of which, that is the Betti number $\beta_0$, counts the number of connected components of a simplicial complex $K$.

Again, consider a connected simplicial complex $K$ and ad vertex $v_0\in K$. An homomorphism from the fundamental group of homotopy to the first homology group 
$$\nu:\pi_1(K,\ZZ)\rightarrow H_1(K,\ZZ)$$
can be obtained {via} the morphism 
$$\sigma:P_n\rightarrow K,$$
where $P_n$ is the $n$-edges closed polygon. In this way, the homomorphism $\nu$ is defined as follows
$$\nu\left([P_n]_{\pi_1}\right)= \sigma_*[P_n]_{\pi_1}:=[\sigma(P_n)]_{H_1},$$ 
where $[P_n]_{\pi_1}$ is the class of $P_n$ in $\pi_1(K,\ZZ)$ and $[\sigma(P_n)]_{H_1}$ is the homology class of $\sigma(P_n)$. It is not difficult to prove that $\nu$ is surjective and its kernel is given by the commutators $[\alpha,\beta]=\alpha*\beta*\alpha^{-1}*\beta^{-1}$, $\alpha,\beta\in\pi_1(K,\ZZ)$. Hence, the homology module $H_1(K,\ZZ)$ represents the classes of loops in the simplicial complex $K$. The same methods can be applied to any connected components of $K$ whenever it is not connected. Therefore, the Betti number $\beta_1$ counts the number of loops
{(one dimensional holes)} that are present in a simplicial complex $K$.
 
Finally, as far as the module $H_2(K,\ZZ)$ is concerned, consider a $1$-connected simplicial complex $K$, i.e. it is connected and simply connected. Then it is possible to represent any classes in $H_2(K,\ZZ)$ by means of the geometric representation $\phi:P\rightarrow K$, where $P$ is a $2$-sphere. Following the idea carried out for the $H_1(K,\ZZ)$, we can say that the homology module $H_2(K,\ZZ)$ characterises the voids inside the simplicial complex $K$.


\section{Statistical models and geometric networks complexity measure}\label{sec3}

We now start describing the geometric entropy that was firstly introduced in \cite{FMP} and further investigated in \cite{Franzosi15}.
\\
\
Consider $n$ \textit{real} random variables (r.v.) $X_1,\ldots,X_n$ with joint probability distribution $p(x;\theta)$ given by the $n$-variate Gaussian density function
\begin{equation}
\label{Gaussian}
p(x;\theta)=\frac{\exp\Big[-\frac{1}{2}\,x^t\,C^{-1}(\theta)\,x\Big]}{\sqrt{(2\pi)^n\,\det\, C(\theta)}},
\end{equation}
which is characterized by $m$ \textit{real} parameters $\theta^1,\ldots,\theta^m$, i.e. the entries of the covariance matrix $C(\theta)$. Here, $t$ is the transposition and $x=(x_1,\ldots,x_n)\in\RR^n$ is the vector of values that $X_1,\ldots,X_N$ take on a sampling space $\Omega$. In addition, we assume mean-values being zero.

Consider now the collection of $n$-variate Gaussian density functions
\begin{equation}
\label{statmodel}
{\cal P}=\{p_\theta= p(x;\theta);\theta\in\Theta\subset\RR^m\},
\end{equation}
where $p(x;\theta)$ is as in Eq. \eqref{Gaussian} and $\Theta:=\{\theta\in\RR^m;C(\theta)>0\}$. So defined, $\cal P$ is an $m$-dimensional statistical model on $\RR^n$. Since a parameter $\theta\in\Theta$ uniquely describes  distribution $p_\theta$, the mapping $\varphi:{\cal P}\rightarrow \RR^m$ defined by $\varphi(p_\theta)=\theta$ is one-to-one. We can thereby consider $\varphi=[\theta^i]$ as a system of local coordinates for $\cal P$. Hence, assuming parametrizations which are $C^\infty$ we can turn $\cal P$ into a differentiable manifold, that is called a \textit{statistical manifold} \cite{amari}. 

Given $\theta\in\Theta$, the Fisher information matrix of $\cal P$ at $\theta$ is the $m\times m$ matrix $G(\theta)=[g_{ij}]$, where the $ij$ entry is defined by
\begin{equation}
\label{gFR}
g_{ij}(\theta):=\int_{\RR^n}dx\,p(x;\theta)\,\partial_i\log p(x;\theta)\partial_j\log p(x;\theta),
\end{equation}
with $\partial_i$ standing for $\frac{\partial}{\partial \theta^i}$. The matrix $G(\theta)$ is symmetric and positive semidefinite \cite{amari}. From here on, we assume it is positive definite; in such a way, $\Theta$ can be endowed with the proper Riemannian metric $g(\theta)=\sum_{i,j=1}^m g_{ij}\ d\theta^i\otimes d\theta^j$ and the manifold ${\cal M}:=(\Theta,g(\theta))$ is a Riemannian manifold.

From Eq. \eqref{Gaussian} the integral in \eqref{gFR} turns out to be a Gaussian one which is easily tractable. If, in addition, we assume that non-diagonal entries $c_{ij}$ of the covariance matrix $C(\theta)$ can take only $0$ or $1$ values, then an explicit analytical relation holds between entries of the matrix $G(\theta)$ and those of the matrix $C(\theta)$ \cite{FMP}, and is given by
\begin{equation}
\label{analytic}
g_{ij}=\frac{1}{2}(c_{ij}^{-1})^2,
\end{equation}
where $c_{ij}^{-1}$ is the $ij$ entry of the inverse of the covariance matrix $C(\theta)$. 

As usual in mathematics, a geometric object is endowed with an over structure in order to employ more efficient tools to describe it (e.g. bundles over manifolds, coverings over topological spaces, and so on). Likewise, we want to endow a network (a discrete system) with a Riemannian manifold (a differentiable and continuous system). Among all the probabilistic methods the random walk one \cite{Noh04} allows a geometric approach through the Green function giving rise to a metric \cite{Mathieu08}. However, our geometric approach is different since, beside the adjacency matrix also the variances of a Gaussian distribution of random variables are taken into account. In fact, we basically exploit two elements: one is the functional relation in Eq. \eqref{analytic} between Fisher matrix $G(\theta)$ and Covariance matrix $C(\theta)$; the other one is the \textit{adjacency matrix} $A$ of a network.

First of all, we interpret random variables $X_1,\ldots,X_n$ as sitting on vertices of a network which is assumed a simple and undirected graph. The \textit{bare} system is assumed as a network without connections among the vertices. In this case we can consider $X_1,\ldots,X_n$ statistically independent. So, their covariance matrix is the $n\times n$ diagonal matrix $C_0(\theta)$ with entries given by
$$
(c_0)_{ii}:=\theta^i=\int_{\RR^n}\,dx\,p(x;\theta)x_i^2,\; i=1,\ldots,n,
$$  
where $p(x;\theta)$ is given by \eqref{Gaussian}.
The parameter space reads as $\Theta_0=\{\theta^i>0;\,i=1,\ldots,n\}$. Furthermore, from Eq. \eqref{analytic} it follows that the Fisher information matrix is the diagonal $n\times n$ matrix $G_0=\frac{1}{2}\left[\left(\frac{1}{\theta^i}\right)^2\right]_{ii}$, for $i=1,\ldots,n$ and $\theta^i\in\Theta_0$. Finally, we can associate to the bare network the statistical Riemannian manifold ${\cal M}=(\Theta_0,g_0)$, with
\begin{equation}
\label{g0}
\Theta_0=\{\theta^i>0;\,i=1,\ldots,n\},\quad g_0=\frac{1}{2}\sum_{i=1}^n \Big(\frac{1}{\theta^i}\Big)^2\,d\theta^i\otimes d\theta^i.
\end{equation}

At this point, in order to take into account the possible connections among vertices of a network, we consider its adjacency matrix $A$. Then, let the map $\psi_{\theta}:A(n,\RR)\rightarrow GL(n,\RR)$ be defined by
\begin{equation}
\label{psi}
\psi_{\theta}(A):=C_0(\theta)+A,
\end{equation}
where $A(n,\RR)$ denotes the set of symmetric $n\times n$ matrices over $\RR$ with vanishing diagonal elements that can represent any simple undirected graph. The manifold associated to the network is the Riemannian manifold $\widetilde{M}=(\widetilde{\Theta},\widetilde{g})$ given by
\begin{equation}
\label{paramvary}
\widetilde{\Theta}:=\{\theta\in\Theta_0;\,\psi_{\theta}(A) \,\mbox{is positive definite}\}
\end{equation}
and $\widetilde{g}=\sum_{ij}\,\widetilde{g}_{ij}\,d\theta^i\otimes d\theta^j$ with components
\begin{equation}
\label{gvary}
\widetilde{g}_{ij}=\frac{1}{2}\Big(\psi_{\theta}(A)_{ij}^{-1}\Big)^2
\end{equation}
where $\psi_{\theta}(A)_{ij}^{-1}$ is $ij$ entry of inverse of the invertible matrix $\psi_{\theta}(A)$.

Hence, given a network with adjacency matrix $A$ and associated manifold $\widetilde{M}=(\widetilde{\Theta},\widetilde{g})$, we put forward the following network complexity measure \cite{Franzosi15},
\begin{equation}
\label{entropy}
\mathcal{S}=\ln{\cal V}(A),
\end{equation}
where ${\cal V}(A)$ is the volume of the manifold $\widetilde{M}$ obtained from the volume element
\begin{equation}
\label{volumeform}
\nu_{\widetilde{g}}=\sqrt{\det\widetilde{g}(\theta)}\,d\theta^1\,\wedge\ldots\wedge\,d\theta^n.
\end{equation}
Unfortunately, in such a way ${\cal V}(A)$ is ill-defined since the parameter space $\widetilde{\Theta}$ is not compact and since $\det\widetilde{g}(\theta)$ diverges as $\det\psi_{\theta}(A)$ approaches to zero for some $\theta^i$. Thus, as it is usual \cite{Leibb75}, we regularize it as follows
\begin{equation}
\label{regular}
{\cal V}(A)=\int_{\widetilde{\Theta}}{\cal R}(\psi_{\theta}(A))\,\nu_{\widetilde{g}},
\end{equation}
where ${\cal R}(\psi_{\theta}(A))$ is any suitable regularizing function. This should be a kind of compactification of the parameter space and with the excision of those sets of $\theta^i$ values which make $\det\widetilde{g}(\theta)$ divergent.

\section{Homology groups and geometric--entropy complexity of networks}\label{sec4}

In this section we apply the methods so far described, i.e. we compute the homology groups of given networks and compare them against the geometric--entropy values of the same networks. Actually, we refer to the dimensions of those groups via the Betti numbers and work out a relation between them and the geometric--entropy. In such a way, we show a clear detection of this kind of network topological features by means of the entropy ${\cal S}$ of the Eq. \eqref{entropy}.

The aim of this work consists of checking to what extent the geometric--entropy ${\cal S}$ can be sensitive to some topological property of networks beyond its effectiveness beforehand checked as a network complexity measure \cite{Franzosi16}.
In order to get well-grounded insights about the relation between the Betti numbers and the values of ${\cal S}$, we firstly consider low--size graphs upon which analytical methods can be worked out. Once that is set up, we investigate large--size networks particularly focusing on two graph ensembles, namely the random graphs and the scale--free networks.

\subsection{Analytical Results}

Let us consider five nodes graphs for our preliminary investigation. The clique graph interpretation thereby ascribes the simplicial complex $K_4$ as a $4$--simplex, i.e. a $5$ nodes fully connected network. In order to show how the entropy ${\cal S}$ works, we refer to two chains within $K_4$ as in Tab. \ref{Tab1}.

\begin{table}
[ht] \caption{The value of $\mathcal{S}$ for networks with five nodes against topological dimension}\label{Tab1}
\vspace{0.2cm}
\begin{tabular}{|cll|cl|cl|}
\hline Network & & & & \mbox{dim} & ${\cal S}$&\\
\hline
\begin{tikzpicture}[thick,scale=0.5]
   
     \draw\foreach \x in {0}
     {
         (\x:2) node{0} -- (\x+72:2) node{1}  (\x+144:2) node{2} (\x+216:2) node{3}  (\x+288:2) node{4}   (\x+360:2) node{0}
           };
 \end{tikzpicture} & & & & 1& $0.4012$&\\
 \hline
\end{tabular}
\begin{tabular}{|cll|cl|cl|}
\hline Network & & & & \mbox{dim} & ${\cal S}$ &\\
\hline
\begin{tikzpicture}[thick,scale=0.5]
   
     \draw[fill=yellow!20]\foreach \x in {0}
     {
         (\x:2) node{0} -- (\x+72:2) node{1} -- (\x+144:2) node{2} (\x+216:2) node{3}  (\x+288:2) node{4}   (\x+360:2) node{0}
(\x+144:2) node{2} -- (\x:2) node{0}};
 \end{tikzpicture} & & & & 2& $0.4179$&\\
 \hline
\end{tabular}
\end{table}

The network on the left side of Tab. \ref{Tab1} is a chain with one $1$-simplex and three $0$-simplexes; from relation \eqref{dimension} it follows that its topological dimension is  $1$. Whereas, the network on the right side of Tab. \ref{Tab1} is a chain with one $2$-simplex and two $0$-simplexes and from \eqref{dimension} we have that its topological dimension is $2$.

According to graph theory \cite{godsil}, the adjacency matrices of these networks are
\begin{equation}
\label{adj1-2}
A_1=\left(\begin{array}{ccccc}
0&1&0&0&0\\
1&0&0&0&0\\
0&0&0&0&0\\
0&0&0&0&0\\
0&0&0&0&0
\end{array}\right),
\quad
A_2=\left(\begin{array}{ccccc}
0&1&1&0&0\\
1&0&1&0&0\\
1&1&0&0&0\\
0&0&0&0&0\\
0&0&0&0&0
\end{array}\right),
\end{equation}
respectively. Consider now $5$ independent random variables $X_1,\ldots,X_5$ sitting on the five completely disconnected nodes of the network,  with Gaussian joint probability distribution given by
\begin{equation}
\label{PxT5}
p(x;\theta)=\frac{\exp\left[-\frac{1}{2} x^t C^{-1}_0(\theta) x\right]}{\sqrt{(2\pi)^5\det C_0(\theta)}},
\end{equation}
where $x\in\RR ^5$ and covariance matrix $C_0(\theta)$ is given by
\begin{equation}
\label{cov0}
C_0=\left(\begin{array}{ccccc}
\theta^1&0&0&0&0\\
0&\theta^2&0&0&0\\
0&0&\theta^3&0&0\\
0&0&0&\theta^4&0\\
0&0&0&0&\theta^5
\end{array}\right),
\quad \theta^i=\int_{\RR^5 } dx\ p(x;\theta)\ x_i^2,\quad \forall i=1,\ldots,5.
\end{equation}

From Eq. \eqref{psi} we obtain that
\begin{equation}
\label{psi1-2}
\psi_{\theta}(A_1)=\left(\begin{array}{ccccc}
\theta^1&1&0&0&0\\
1&\theta^2&0&0&0\\
0&0&\theta^3&0&0\\
0&0&0&\theta^4&0\\
0&0&0&0&\theta^5
\end{array}\right),
\quad
\psi_{\theta}(A_2)=\left(\begin{array}{ccccc}
\theta^1&1&1&0&0\\
1&\theta^2&1&0&0\\
1&1&\theta^3&0&0\\
0&0&0&\theta^4&0\\
0&0&0&0&\theta^5
\end{array}\right).
\end{equation}
In order to associate a parameter space to each network, the matrices $\psi_{\theta}(A_1)$ and $\psi_{\theta}(A_2)$ must be positive definite, as indicated by Eq. \eqref{paramvary}. Such a condition is fulfilled by imposing that each main minor of the matrix has positive determinant. Thereby, we arrive at
\begin{eqnarray}
\label{parvary1-2}
&&\widetilde{\Theta}_1=\left\{\theta=(\theta^1,\ldots,\theta^5); \theta^1>0,\theta^1\theta^2>1,\theta^3>0,\theta^4>0,\theta^5>0\right\}\nonumber\\
\nonumber\\
&&\widetilde{\Theta}_2=\left\{\theta=(\theta^1,\ldots,\theta^5); \theta^1>0,\theta^1\theta^2>1,\theta^3>\frac{\theta^1+\theta^2-2}{\theta^1\theta^2-1},\theta^4>0,\theta^5>0\right\}.
\end{eqnarray}
Moreover, from Eq. \eqref{gvary} we endow $\widetilde{\Theta}_1$ and $\widetilde{\Theta}_2$, defined in \eqref{parvary1-2}, with Riemannian metrics, hence obtaining ${\cal M}_1=(\widetilde{\Theta}_1,\widetilde{g}_1) $ and ${\cal M}_2=(\widetilde{\Theta}_2,\widetilde{g}_2) $ as manifolds associated to the networks. Here, $\widetilde{g}_i$ and $\widetilde{g}_2$ are given by
\begin{eqnarray}
\label{gvary1-2}
&&\widetilde{g}_1=\frac{\left(\theta^2\right)^2d\theta^1\otimes d\theta^1+\left(\theta^1\right)^2d\theta^2\otimes d\theta^2}{2\left(\theta^1\theta^2-1\right)^2}+\frac{d\theta^4\otimes d\theta^4}{2\left(\theta^4\right)^2}+\frac{d\theta^5\otimes d\theta^5}{2\left(\theta^5\right)^2}+\frac{d\theta^1\otimes d\theta^2}{\left(\theta^1\theta^2-1\right)^2}\nonumber\\
\nonumber\\
&&\widetilde{g}_2=\frac{\left(\theta^2\theta^3-1\right)^2d\theta^1\otimes d\theta^1+\left(\theta^1\theta^3-1\right)^2 d\theta^2\otimes d\theta^2+\left(\theta^1\theta^2-1\right)^2 d\theta^3\otimes d\theta^3}{2\left(\theta^1(\theta^2\theta^3-1\right)-\theta^2-\theta^3+2)^2}\nonumber\\
&&+\frac{d\theta^4\otimes d\theta^4}{2\left(\theta^4\right)^2}+\frac{d\theta^5\otimes d\theta^5}{2\left(\theta^5\right)^2}+\frac{(1-\theta^3)^2d\theta^1\otimes d\theta^2+(1-\theta^2)^2d\theta^1\otimes d\theta^3}{\left(\theta^1(\theta^2\theta^3-1\right)-\theta^2-\theta^3+2)^2}.
\end{eqnarray}
The next ingredients necessary to compute the geometric--entropy values against the two networks are the volume elements computed through Eq. \eqref{volumeform}. However, as we can see from the analytical expressions for $\det\widetilde{g}_1$ and $\det\widetilde{g}_2$, 
\begin{eqnarray}\label{det1-2}
&&\det\widetilde{g}_1=\frac{1+\theta^1\theta^2}{32\left(\theta^3\theta^4\theta^5\right)^2\left(\theta^1\theta^2-1\right)^3}\nonumber\\
\nonumber\\
&&\det\widetilde{g}_2=\frac{\left(\theta^1\right)^2\left(\left(\theta^2\theta^3\right)^2-1\right)+2\theta^1\left(\theta^2+\theta^3-2\theta^2\theta^3\right)^2-\left(\theta^2-\theta^3\right)^2}{32\left(\theta^4\theta^5\left(\theta^1+\theta^2+\theta^3-\theta^1\theta^2\theta^3-2\right)^2\right)^2},
\end{eqnarray}
it turns out that both ${\cal V}(A_1)$ and ${\cal V}(A_2)$ are ill-defined. In fact, on the one side, the numerator of $\det\widetilde{g}_i(i=1,2)$ diverges as $\theta^j$ goes to infinity for some $j\in\{1,\ldots,5\}$; on the other side, the denominator of  $\det\widetilde{g}_i$ goes to zero as $\theta^j$ approaches the lower bound of $\widetilde{\Theta}_i(i=1,2)$ for some $j\in\{1,\ldots,5\}$.

In order to overcome this difficulty,  we introduce a regularizing function ${\cal R}(\psi_{\theta}(A_{i}))$. This issue has been widely tackled in the literature (see for instance Ref. \cite{Leibb75}). As far as our approach is concerned, the aim is to make meaningful the following integrals
\begin{equation}\label{entropy1-2}
\int_{\widetilde{\Theta}_1}\ {\cal R}(\psi_{\theta}(A_{1}))\ \nu_{\widetilde{g}_1},\quad \int_{\widetilde{\Theta}_2}\ {\cal R}(\psi_{\theta}(A_{2}))\ \nu_{\widetilde{g}_2},
\end{equation}
where $\nu_{\widetilde{g}_1}=\sqrt{\det\widetilde{g}_1}\ d\theta^1\wedge\ldots\wedge d\theta^5$ and $\nu_{\widetilde{g}_2}=\sqrt{\det\widetilde{g}_2}\ d\theta^1\wedge\ldots\wedge d\theta^5$. Supported by a general result (see Corollary 2 in \cite{Felice17}) we consider the following regularizing function,
\begin{equation}\label{regf}
{\cal R}(\psi_{\theta}(A_{i})):=\left(H(h-\mbox{Tr}(C_0))+H(\mbox{Tr}(C_0)-h)\ e^{-\mbox{Tr}(C_0)}\right)\log\left(1+\left(\det\psi_{\theta}(A_{i})\right)^5\right).
\end{equation}
Here, $H(\cdot)$ denotes the Heaviside step function and $\mbox{Tr}$ is the trace operator, while $h$ is a \textit{real} number accounting for a finite range of $\theta^j$s values where integrals in \eqref{entropy1-2} are substantial. Regularizing function in \eqref{regf} depends on structure of networks through adjacency matrix $A_i$ and cures all the possible divergences of integrals \eqref{entropy1-2}, due to the functional forms in \eqref{det1-2}. Indeed, if $\theta^j$ goes to infinity for some $j=1,\ldots,5$, then the integrand is killed to zero by function $\left(H(h-\mbox{Tr}(C_0))+H(\mbox{Tr}(C_0)-h)\ e^{-\mbox{Tr}(C_0)}\right)$. Moreover, if $\det\psi_{\theta}(A_{i})$ goes to zero then the singularity is removed by the $\log\left(1+\left(\det\psi_{\theta}(A_{i})\right)^5\right)$ exploiting the well-known relation $\lim_{x\rightarrow 0}\frac{\log(1+x)}{x}=1$.

We are now ready to compute the values of the integrals in \eqref{entropy1-2} through the regularizing functions as in \eqref{regf} which are shown in Tab. \ref{Tab1}. The difference of entropic values between the $1$-dimensional chain and the $2$-dimensional chain is a particular result of a more general one \cite{FMP}. In addition, it is worth remarking that integrals in \eqref{entropy1-2} do not depend on the particular label permutation of the network  \cite{FMP}. Hence, if two networks are isomorphic, then the values of ${\cal S}$ as in \eqref{entropy} coincide.

Actually, two isomorphic networks, intended as clique graphs, have the same topological features; thereby, they have the same homotopy and homology groups. As a consequence, whenever the Betti numbers $\beta_0$ and $\beta_1$ are different, then the networks are not isomorphic. In Tab. \ref{Tab2} we show the geometric--entropic values of ${\cal S}$ as in \eqref{entropy} computed for several $5$--nodes networks. In particular, we point out that ${\cal S}$ reflects the differences of their first homology groups. Indeed, by following the values on $\beta_0$ in Tab. \ref{Tab2}, we can  see that the larger the number of connected components of a network, the lower the value of the entropy \eqref{entropy}. On the contrary, according to the values of $\beta_1$ in Tab. \ref{Tab2}, the larger the number of cycles the more complex the network. In addition, cycles mostly influence the complexity with respect to topological dimension as it follows by comparing the values of the entropy \eqref{entropy} of the last two networks on the right side of Tab. \ref{Tab2}. 

\begin{table}
[ht] \caption{The value  of $\mathcal{S}$ for networks with five nodes against topological dimension and Betti numbers $\beta_0$ and $\beta_1$}  \label{Tab2}
\vspace{0.2cm}
\begin{tabular}{|clll|c|c|c|c|}
\hline Network & & & & \mbox{dim}& $\beta_0$ & $\beta_1$ & $\mathcal{S}$\\
\hline
\begin{tikzpicture}[thick,scale=0.5]
   
     \draw\foreach \x in {0}
     {
         (\x:2) node{0} -- (\x+72:2) node{1}  (\x+144:2) node{2} (\x+216:2) node{3}  (\x+288:2) node{4}   (\x+360:2) node{0}
           };
 \end{tikzpicture} & & & & $1$& $4$ & $0$ & $0.4012$\\
\hline
\begin{tikzpicture}[thick,scale=0.5]
   
     \draw\foreach \x in {0}
     {
         (\x:2) node{0} -- (\x+72:2) node{1}  (\x+144:2) node{2} (\x+216:2) node{3}  (\x+288:2) node{4} --  (\x+360:2)  node{0}
           };
 \end{tikzpicture} & & & & $1$& $3$ & $0$ & $1.5258$\\
 \hline
 \begin{tikzpicture}[thick,scale=0.5]
    
      \draw\foreach \x in {0}
      {
          (\x:2) node{0} -- (\x+72:2) node{1}  (\x+144:2) node{2} -- (\x+216:2) node{3}  (\x+288:2) node{4} --  (\x+360:2)  node{0}
            };
  \end{tikzpicture} & & & & $1$& $2$ & $0$ & $1.7153$\\
  \hline
   \begin{tikzpicture}[thick,scale=0.5]
      
        \draw\foreach \x in {0}
        {
            (\x:2) node{0} -- (\x+72:2) node{1} -- (\x+144:2) node{2} -- (\x+216:2) node{3}  (\x+288:2) node{4}   (\x+360:2)  node{0}
              };
    \end{tikzpicture} & & & & $1$& $2$ & $0$ &$2.0849$\\
    \hline
    \begin{tikzpicture}[thick,scale=0.5]
          
            \draw\foreach \x in {0}
            {
                (\x:2) node{0} -- (\x+72:2) node{1} -- (\x+144:2) node{2} -- (\x+216:2) node{3} -- (\x+288:2) node{4}   (\x+360:2)  node{0}
                  };
        \end{tikzpicture} & & & &$1$& $1$ & $0$ & $2.8739$\\
        \hline
        \end{tabular}
        \begin{tabular}{|clll|c|c|c|c|}
        \hline Network & & & & \mbox{dim}& $\beta_0$  & $\beta_1$ & $\mathcal{S}$\\
        \hline
        \begin{tikzpicture}[thick,scale=0.5]
                  
                    \draw\foreach \x in {0}
                    {
                        (\x:2) node{0} -- (\x+72:2) node{1} -- (\x+144:2) node{2} -- (\x+216:2) node{3}  (\x+288:2) node{4}   (\x+360:2)  node{0}
                   (\x:2) node{0} --  (\x+216:2) node{3}};
                \end{tikzpicture} & & & &$1$& $2$ & $1$ & $3.0419$\\
                \hline
                 \begin{tikzpicture}[thick,scale=0.5]
                          
                            \draw\foreach \x in {0}
                            {
                                (\x:2) node{0} -- (\x+72:2) node{1} -- (\x+144:2) node{2} -- (\x+216:2) node{3} -- (\x+288:2) node{4}  -- (\x+360:2)  node{0}
                                  };
                        \end{tikzpicture} & & & &$1$& $1$ & $1$ & $3.2699$\\
                        \hline
                          \hline
                            \begin{tikzpicture}[thick,scale=0.5]
                                   \draw[fill=yellow!20] \foreach \x in {0}
                                  {
                                  (\x:2) node{0} -- (\x+72:2) node{1} -- (\x+144:2) node{2}  (\x+288:2) node{3} -- (\x+216:2) node{4}   (\x+360:2)  node{0}
                                  (\x+144:2) node{2} -- (\x:2) node{0}
                                  (\x+72:2) node{1}  (\x+288:2) node{3}
                                 (\x+144:2) node{2} -- (\x+216:2) node{4}
                                 (\x+288:2) node{3} -- (\x:2) node{0} };
                                  \end{tikzpicture} & & & &$2$& $1$ & $1$ & $4.7617$\\
       \hline
  \begin{tikzpicture}[thick,scale=0.5]
         \draw[fill=yellow!20] \foreach \x in {0}
        {
        (\x:2) node{0} -- (\x+72:2) node{1} -- (\x+144:2) node{2} -- (\x+252:1) node{3}  (\x+252:2) node{4}  -- (\x+360:2)  node{0}
        (\x+72:2) node{1} -- (\x+252:1) node{3}
       (\x+144:2) node{2} -- (\x+252:2) node{4}
       (\x+252:1) node{3} -- (\x:2) node{0} };
        \end{tikzpicture} & & & &$2$& $1$ & $1$ & $5.4758$\\
                        \hline
                        \begin{tikzpicture}[thick,scale=0.5]
  \draw\foreach \x in {0}
 {
 (\x:2) node{0} -- (\x+72:2) node{1} -- (\x+144:2) node{2} -- (\x+45:0) node{3}  (\x+252:2) node{4}  -- (\x+360:2)  node{0}
(\x+144:2) node{2} -- (\x+252:2) node{4}
(\x+45:0) node{3} -- (\x:2) node{0} };
 \end{tikzpicture} & & & &$1$& $1$ & $2$ & $6.0841$\\
 \hline
\end{tabular}
\end{table}

\bigskip

In Tab. \ref{Tab3} we can see that networks with the same Betti numbers $\beta_0$ and $\beta_1$ do not have the same values of ${\cal S}$. This is in agreement with results in \cite{Franzosi16} where it has been shown that networks with different degree distributions do not have the same complexity computed by ${\cal S}$. Hence, characterizing networks complexity through their first homology groups is not enough. Beyond these topological features, it is necessary to consider also other aspects as it is well--known in the vast literature around. For this reason, the fact that ${\cal S}$ takes different values for networks having the same Betti numbers $\beta_0$ and $\beta_1$ is not a validity restriction of  the geometric entropy.

\begin{table}
[ht] \caption{The value of $\mathcal{S}$ for networks with five nodes and same homology groups}\label{Tab3}
\vspace{0.2cm}
\begin{tabular}{|clll|c|c|c|c|}
\hline Network & & & & \mbox{dim}& $\beta_0$ & $\beta_1$ & $\mathcal{S}$\\
\hline
 \begin{tikzpicture}[thick,scale=0.5]
    
      \draw\foreach \x in {0}
      {
          (\x:2) node{0} -- (\x+72:2) node{1}  (\x+144:2) node{2} -- (\x+216:2) node{3}  (\x+288:2) node{4} --  (\x+360:2)  node{0}
            };
  \end{tikzpicture} & & & & $1$& $2$ & $0$ & $1.7153$\\
 \hline
\end{tabular}
\begin{tabular}{|clll|c|c|c|c|}
\hline Network & & & & \mbox{dim}& $\beta_0$ & $\beta_1$ & $\mathcal{S}$\\
\hline
\begin{tikzpicture}[thick,scale=0.5]
      
        \draw\foreach \x in {0}
        {
            (\x:2) node{0} -- (\x+72:2) node{1} -- (\x+144:2) node{2} -- (\x+216:2) node{3}  (\x+288:2) node{4}   (\x+360:2)  node{0}
              };
    \end{tikzpicture} & & & & $1$& $2$ & $0$ &$2.0849$\\
 \hline
\end{tabular}
\end{table} 

\subsection{Numerical Results}

We now investigate the behaviour of the geometric--entropy ${\cal S}$, given in \eqref{entropy}, by comparing it with respect to the behaviour of the Betti numbers computed for two ensembles of networks \cite{Bianconi}, namely the random graphs and the scale--free networks. 

The numerical computation of ${\cal S}$ has been performed on $200$--nodes networks by exploiting the functional form \eqref{gvary} of the components of the varied Fisher--Rao metric $\widetilde{g}$. Then, the volume regularization has been obtained at first by restricting the manifold support ${\widetilde{\Theta}}\subset \mathbb{R}^n$ to an hypercube, and we worked out Monte Carlo estimates of the average 
$
\left\langle\sqrt{\det \widetilde{g}} \right\rangle =\int \sqrt{\det \widetilde{g}}\;d\theta^1\wedge\ldots\wedge d\theta^n/\int \;d\theta^1\wedge\ldots\wedge d\theta^n
$
by generating Markov chains inside ${\widetilde{\Theta}}$. In this case, the number of random configurations considered varies between $10^4$ and $10^6$. Finally, the regularization procedure of the volume is obtained by excluding those points where the value of $\sqrt{\det \widetilde{g}}$ exceeds $10^{308}$ (the numerical overflow limit of the computers used) \cite{Franzosi15}. 

One of the basic models of random graphs is the uniform random graph $\mathbb{G}(n,k)$. This is devised by choosing with uniform probability a graph from the set of all the graphs having $n$ vertices and $k$ edges, with $k$ a nonnegative integer \cite{Lucz}. Scale--free networks can be obtained as special cases of random graphs with a given degree distribution showing thereof a power--law degree distribution. These are described by two parameters, $\alpha$ and $\gamma$ , which define the size and the density of a network; hence, given the number of nodes $n$ with degree $d$, these models, denoted $\mathbb{G}_{\alpha,\gamma}$, assigns a uniform probability to all graphs with $n=e^{\alpha} d^{-\gamma}$ \cite{Boccaletti06}. 

\begin{figure}[ht]\centering
\includegraphics[scale=.75]{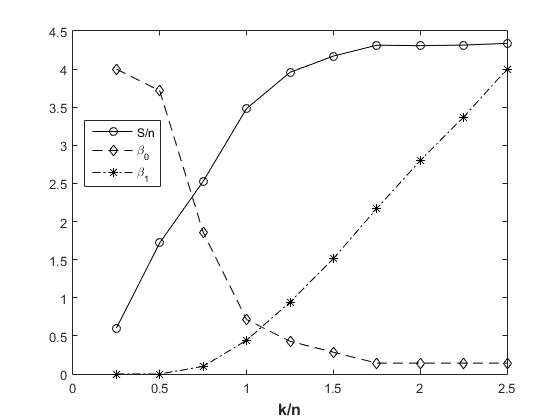}
\caption[10 pt]{Behaviours of the geometric--entropy ${\cal S}/n$ and the Betti numbers $\beta_0$, $\beta_1$ of $\mathbb{G}(n,k)$ as functions of the number $k$ of randomly chosen links of weights equal to $1$.}\label{RG}
\end{figure}

In Fig. \ref{RG} we report the behaviour of ${\cal S}/n$ of $\mathbb{G}(n,k)$ vs $k/n$ for a fixed value of $n$, namely $n=200$, together with the behaviour of the Betti numbers $\beta_0$ and $\beta_1$ of $\mathbb{G}(n,k)$ \cite{Vittorio}. By interpreting $\mathbb{G}(n,k)$ as clique graphs, we can see a perfect correlation between ${\cal S}/n$ and the Betti number $\beta_0$. This is not surprising as $\beta_0$ reflects the number of connected components of $\mathbb{G}(n,k)$ and the appearance of a giant component is accounted for $\mathbb{G}(n,k)$ when $k/n>0.5$. Whereas the correlation of ${\cal S}$ with the Betti number $\beta_1$ of $\mathbb{G}(n,k)$ is subtler to interpret. Indeed, $\beta_1$ rapidly increases its value as $k/n$ increases, contrary to ${\cal S}/n$ which shows a saturation when the numbers of $k$ is large. Actually, $\beta_1$ counts the cycles of $\mathbb{G}(n,k)$ and if on one side $\beta_1$ equals zero when $k\leq n/2$ reflecting a well--known theoretical result \cite{Lucz}, on the other side it is not enough for completely describing the topology of $\mathbb{G}(n,k)$ when $k$ is larger than $5/2\ n $. However, the geometric--entropy ${\cal S}$ properly correlates with the Betti numbers $\beta_0$ and $\beta_1$ when $k\leq 1.5\ n$ and the only homology groups of $\mathbb{G}(n,k)$ are $H_0$ and $H_1$.

The scale--free network ensemble is recovered in the case of networks with a finite number of cycles \cite{Bianconi07}. In order to highlight the connection of the geometric--entropy ${\cal S}$ of Eq. \eqref{entropy} with the topology of the power--law random graph model $\mathbb{G}_{\alpha,\gamma}$ intended as a clique graph we have proceeded as follows. We considered networks of $n = 200$ nodes for which, without loss of generality, we set $\alpha=0$. For each value of $\gamma$ , we selected $10$ different realisations of the networks,
each realisation having the same value of $k/n$. Actually, because of the practical difficulty of getting different
realisations of a scale--free network with exactly the same value of $k/n$ at different $\gamma$ values, we accepted a spread of values in the range $0.7-0.85$.

\begin{figure}[ht]\centering
\includegraphics[scale=.50]{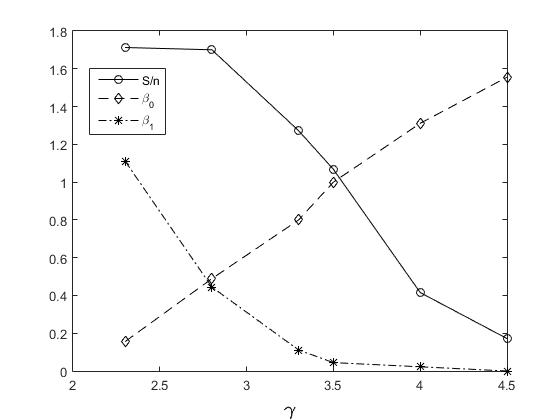}
\caption[10 pt]{Behaviours of the entropy ${\cal S}/n$ and Betti numbers $\beta_0$, $\beta_1$ of power-law $\mathbb{G}_{0,\gamma}(200,k)$ networks as a function of the exponent $\gamma$; $k$ values varied - according
to the realization of the RG and, independently, of $\gamma$ - approximately in the range $160-500$.}\label{SF}
\end{figure}

In Fig. \ref{SF} we report the behaviour of the geometric--entropy ${\cal S}/n$ of the power--law random graphs $\mathbb{G}_{\alpha,\gamma}(n,k)$ when $\alpha=0$, $n=200$, and the exponent $\gamma$ is in the range $2.3 < \gamma < 4.5$ together with the behaviours of the Betti numbers of the same networks $\mathbb{G}_{\alpha,\gamma}(n,k)$ \cite{Vittorio}. The pattern of ${\cal S}/n$ displays a clear correlation with the Betti number $\beta_1$. In addition, in the range of $\gamma$ considered here, the only topological features of $\mathbb{G}_{\alpha,\gamma}(n,k)$ consist of the homology groups $H_0$ and $H_1$. Hence, a tight correlation between the geometric--entropy ${\cal S}$ and the topology of the power--law random graphs $\mathbb{G}_{\alpha,\gamma}(n,k)$ is found to exist. Indeed, for $2.3 < \gamma < 4.5$ the Euler--Poincar\'e characteristic $\chi(\mathbb{G}_{\alpha,\gamma}(n,k))$ is supplied only by $\beta_0$ and $\beta_1$ as follows from Fig. \ref{SF} and Eq. \eqref{E-P}. 

Finally, in both figures, Fig. \ref{RG} and Fig. \ref{SF}, the pattern of ${\cal S}$ shows a sort of saturation of its values. The reason of this relies on the numerical methods for regularizing the volume computed through the volume form \eqref{volumeform}. Interestingly, the apical values of $\mathbb{G}_{\alpha,\gamma}(n,k)$ and $\mathbb{G}(n,k)$ are different thus allowing to characterise different network ensembles via the geometric--entropy ${\cal S}$ defined in Eq. \eqref{entropy}.

\section{Concluding remarks}\label{sec5}

In this work we have pursued our investigation about the potential applications of a recently defined geometric--entropy which has been shown to perform very well in quantifying networks complexity \cite{FMP,Franzosi15,Franzosi16}. The present investigation focused on the possible use of the mentioned geometric entropy to catch some topological property of networks. By considering networks as clique graphs, we proceeded in a bottom--up analysis. To begin with, small--size networks have been considered and analytical computations have been applied. Then, numerical computations has been used to tackle large--size networks.

The entropy of a network is obtained after having associated to it - on the basis of a probabilistic approach - a Riemannian manifold, and then by computing the volume of this manifold. Since in general this manifold is not compact, we introduced an ``infra--red'' and ``ultra--violet'' regularising function to compactify it. This procedure is independent of networks topology in as much as it is defined up to graph isomorphisms. 

Small--size networks are ascribed to a $4$--dimensional simplicial complex $K$. The analytical computation of ${\cal S}$ for these networks displays a monotonic behaviour of ${\cal S}$ with respect to the Betti numbers $\beta_0$ and $\beta_1$. However, the information about network complexity retained by the geometric entropy cannot be reduced to the only knowledge of the topological properties described by $H_0$ and $H_1$. This is due to the fact that $\beta_0$ and $\beta_1$ do not exhaustively account for the network connectivity. This explains why different values of ${\cal S}$ are found for networks with same $\beta_0$ or $\beta_1$ (see Tab. \ref{Tab3}). Then, passing to a ``coarse--grained'' description for large networks, the connection between ${\cal S}$ and topology becomes clearer.

We have considered two different network ensembles in tackling large--size graphs, that is, the random graphs and the scale--free networks. The entropy of random graphs perfectly correlates with their $\beta_0$. This is not very surprising because $\beta_0$ counts the number of connected components of a graph. Thus, due to the appearance of a giant component in random graphs, $\beta_0$ is expected to asymptotically reach the value $1$, as well ${\cal S}$ saturates for large values of $k$. As far as $\beta_1$ is concerned, we found a proper correlation between the entropy of scale--free networks and their number of cycles. Indeed, scale--free models account for networks with a finite numbers of cycles and the behaviour of ${\cal S}$ properly agrees with the pattern displayed by the Betti number $\beta_1$ of the considered power--law random graphs. This suggests a strong correlation  between ${\cal S}$ and the topology of scale--free networks.

Finally, from both  Figure \ref{RG} and Figure \ref{SF}, we can see that the entropy ${\cal S}$ saturates for some values of $k$ and $\gamma$, respectively. Since these apical values are different, it seems that the geometric--entropy ${\cal S}$ is able to characterise some difference within the network ensemble. This paves the way to further and deeper studies also about this issue.

\begin{acknowledgments}
We are indebted with M. Piangerelli, M. Quadrini, and V. Cipriani of the Computer Science Division of the University of Camerino for computational help.  D.F. also thanks E. Andreotti for useful discussions.
\end{acknowledgments}

\end{document}